\documentclass[%
 pdftex,
 10pt,
 letterpaper,
 prd,
 floatfix,
 twocolumn,
 superscriptaddress,
 nofootinbib,
 amsfonts,
 amssymb,
 amsmath
]{revtex4-2}

\usepackage{graphicx}
\usepackage{cmap} 
\usepackage{ifpdf}
 \ifpdf
 \usepackage{cmap} 
 \usepackage{color,graphicx}
 \usepackage{epstopdf} 
 \else
 \usepackage{color,graphicx}
 \fi
\usepackage{bm}
\usepackage[dvipsnames]{xcolor}
\usepackage{hyperref}
\hypersetup{%
colorlinks=true,%
citecolor=Blue,%
filecolor=Blue,
linkcolor=Blue,
urlcolor=Blue
}
\usepackage[activate={true,nocompatibility},final,tracking=true,kerning=true,spacing=true,factor=1100,stretch=10,shrink=10]{microtype}

\graphicspath{{./}}

\newcommand{\mara}{{MARATHON}}

\newcommand{\ud}  {\mathrm{d}}

\newcommand{\mev} {\ \mathrm{MeV}}
\newcommand{\gevsq}{\ \mathrm{GeV}^2}

\newcommand{\ceps}{\varepsilon}

\newcommand{\eq}[1]{Eq.~(\ref{#1})}

\newcommand{\htri}{{}^3\text{H}}
\newcommand{\hetri}{{}^3\text{He}}

\AtBeginDocument{%
\newwrite\bibnotes
\def\bibnotesext{Notes.bib}
\immediate\openout\bibnotes=\jobname\bibnotesext
\immediate\write\bibnotes{@CONTROL{REVTEX42Control}}
\immediate\write\bibnotes{@CONTROL{%
apsrev42Control,author="08",editor="0",pages="0",title="",year="1"}}
\if@filesw
 \immediate\write\@auxout{\string\citation{apsrev42Control}}%
\fi
}

\allowdisplaybreaks

\begin{document}

\title{Off-shell effects in bound nucleons and parton distributions from\\$^1$H, $^2$H, $^3$H and $^3$He data}

\author{S.~I.~Alekhin}
\affiliation{II. Institut f\"ur Theoretische Physik, Universit\"at Hamburg,
D--22761 Hamburg, Germany}
\author{S.~A.~Kulagin}
\affiliation{Institute for Nuclear Research of the Russian Academy of Sciences,
117312 Moscow, Russia}
\author{R.~Petti}
\affiliation{Department of Physics and Astronomy,
University of South Carolina, Columbia, South Carolina 29208, USA}


\begin{abstract}
\noindent
We report the results of a new global QCD analysis
including deep-inelastic scattering data off $^1$H, $^2$H, $^3$H, and $^3$He targets.
Nuclear corrections are treated in terms of a nuclear convolution approach with off-shell bound nucleons.
The off-shell (OS) corrections responsible for the modification of the structure functions (SFs) of bound nucleons are constrained in a global fit along with the proton parton distribution functions (PDFs) and the higher-twist (HT) terms.
We investigate the proton-neutron difference for the OS correction
and discuss our predictions for the SF ratio $F_2^n/F_2^p$ and the corresponding PDF ratio $d/u$ in the proton,
as well as their correlations with the underlying treatment of the HT terms and of the OS corrections.
In particular, we find that the recent \mara{} data are consistent with equal relative OS corrections for both the proton and the neutron.
\end{abstract}

\maketitle

\section{Introduction}

The parton distribution functions (PDFs) are universal process-independent characteristics of hadrons driving the cross sections of various leptonic and hadronic
processes at high momentum transfer.
The PDFs are usually extracted from global QCD analyses of experimental data at high values of momentum transfer (for a review see, e.g.,~\cite{Accardi:2016ndt,Gao:2017yyd}).
Precise determinations of
PDFs are increasingly important in a variety of tests of the Standard Model and new physics searches.
Nuclear deep-inelastic scattering (DIS) data can be helpful in this context for various reasons.
For instance, the use of nuclear targets with a different proton-neutron content allows one to better constrain the $d$-quark distribution in the proton.
Furthermore, including nuclear DIS data in a QCD analysis improves the statistical significance of the fit.
However, employing those data requires understanding of nuclear effects.

Nuclear effects are usually treated empirically in PDF analyses, employing simple parametrizations of the $A$ and $x$ dependencies (for a review see, e.g.,~\cite{Ethier:2020way,Kovarik:2019xvh}).
Alternatively, we can follow a different approach and employ a microscopic model
accounting for a number of nuclear effects
caused by the energy-momentum distribution of bound nucleons, 
the off-shell (OS) corrections to the nucleon structure functions (SFs),
and meson-exchange currents,
as well as the nuclear propagation of quark-gluon states resulting in the nuclear shadowing effect~\cite{Kulagin:2004ie}.
A number of dedicated studies~\cite{Kulagin:2004ie,Kulagin:2010gd,Kulagin:2007ju,Kulagin:2014vsa,Ru:2016wfx}
indicate that this approach describes with good accuracy the observed nuclear effects in the charged-lepton and neutrino DIS and in the Drell-Yan process, as well as in the $W/Z$ boson production in proton-lead collisions.

In this Letter we report the results of a global QCD analysis, in which we simultaneously
constrain the proton PDFs together with the higher-twist (HT) terms and the OS functions of the nucleon SFs.
To this end, we use the deuterium DIS cross section data from various experiments,
together with recent precision data on the $^3$He to $^3$H ratio of the DIS cross sections from the \mara{} experiment~\cite{MARATHON:2021vqu}.
We study the interplay between the $d/u$ PDF ratio (and the related SF ratio $F_2^n/F_2^p$), the underlying model of HT terms, and the OS corrections.
In particular, in this way, we constrain the proton-neutron asymmetry in the OS corrections to the  SFs.

\section{Theory framework}

For the spin-independent charged-lepton inelastic scattering,
the cross sections are fully described in terms of two SFs, $F_T=2xF_1$ and $F_2$.
In the DIS region of high invariant momentum transfer squared $Q^2$,
in the massless limit,
SFs can be treated in terms of a power series in $Q^{-2}$ (twist expansion) within the operator product expansion (OPE).
The leading twist (LT) SFs are given by a convolution of PDFs with the functions describing the quark-gluon interaction at the scale $Q$, which can be computed perturbatively as a series in the strong coupling constant (see, e.g.,~\cite{Accardi:2016ndt,Gao:2017yyd}).
A finite target mass produces a correction that can be treated within the OPE~\cite{Georgi:1976ve}.
We can then write
\begin{equation}\label{eq:sf}
F_i = F_i^{\text{TMC}} + H_i/ Q^2 + \cdots, 
\end{equation}
where $i=T,2$ and $F_i^{\text{TMC}}$ are the corresponding LT SFs with the account of the target mass correction (TMC)~\cite{Georgi:1976ve},
and $H_i$ describe the twist-4 contribution.
In this study, we consider two commonly used HT models:
(1) additive HT model (aHT) motivated by the OPE, in which we assume $H_i=H_i(x)$  and
(2) multiplicative HT model (mHT)~\cite{Virchaux:1991jc},
in which $H_i$ is assumed to be proportional to the corresponding LT SF, $H_i=F_i^\text{LT}(x,Q^2) h_i(x)$.

To address the nuclear corrections in DIS, we consider this process in the target rest frame and treat it as incoherent scattering off bound nucleons.
The nuclear SFs $F_i^A$ can then be calculated in terms of the bound proton and neutron SFs integrated with
the corresponding spectral functions $\mathcal P^p_A$ and $\mathcal P^n_A$~\cite{Akulinichev:1985ij,Kulagin:1989mu,Kulagin:1994fz,Kulagin:2004ie,Kulagin:2010gd},
\begin{align}
\label{eq:IA}
F_i^A = \int \ud^4 p K_{ij} \left(
\mathcal{P}^p_A F_j^p + \mathcal{P}^n_A F_j^n\right),
\end{align}
where $i,j=T,2$, we assume a summation over the repeated index $j$,
and $K_{ij}$ are the kinematic factors~\cite{Kulagin:2004ie,Alekhin:2022tip}.
The integration is performed over the bound nucleon four-momentum $p$.
The OS nucleon SFs depend on the scaling variable $x'=Q^2/2p\cdot q$,
the DIS scale $Q^2$, and the nucleon invariant mass squared $p^2=p_0^2-\bm p^2$.
This latter dependence originates from both the power TMC terms of the order $p^2/Q^2$ and the OS dependence of the LT SFs.
Following Refs.~\cite{Kulagin:1994fz,Kulagin:2004ie},
we treat the OS correction in the vicinity of the mass shell $p^2=M^2$ by expanding SFs in a power series in $v=(p^2-M^2)/M^2$.
To the leading order in $v$ we have
\begin{align}
\label{SF:OS}
F_i^\text{LT}(x,Q^2,p^2) &=
F_i^\text{LT}(x,Q^2,M^2)\left[ 1+\delta f_i\,v \right],
\\
\label{eq:deltaf}
\delta f_i &= \partial \ln F_i^\text{LT}(x,Q^2,p^2)/\partial \ln p^2,
\end{align}
where the derivative in \eq{eq:deltaf} is taken on the mass shell $p^2=M^2$.
We assume identical functions in \eq{eq:deltaf} for $F_T$ and $F_2$
based on the observation that $F_T\approx F_2$ at large $x$ values,
for which the OS effect is numerically important~\cite{Kulagin:2004ie,Kulagin:2010gd,Alekhin:2017fpf,Alekhin:2022tip}.
We thus suppress the index $i=T,2$ for the function $\delta f$.

The proton (neutron) spectral function $\mathcal P^{p(n)}_A(\ceps,\bm p)$ describes the corresponding energy ($\ceps=p_0-M$) and momentum ($\bm p$) distribution in the considered nucleus at rest.
This function is normalized to the proton (neutron) number.
For the deuteron, the function $\mathcal P^p_d=\mathcal P^n_d$
is fully determined by the deuteron wave function as discussed in detail in Refs.~\cite{Kulagin:2004ie,Alekhin:2022tip}.
For the proton spectral function of $^3$He, $\mathcal P^p_{\hetri}$,
the relevant contributions come from two-body $pn$ intermediate states.
They can be divided into two terms:
the bound $pn$ state, i.e. the deuteron, and the $pn$ states in the  continuum.
The neutron spectral function of $^3$He, $\mathcal P^n_{\hetri}$, involves only the $pp$ continuum states. We then have~\cite{Kulagin:2010gd,Kulagin:2008fm}:
\begin{align}
\mathcal P^p_{\hetri} &= f^d_{\hetri}(\bm p)\delta(E+\ceps_{32}-\ceps_d) + f^{pn}_{\hetri}(E,\bm p),
\label{eq:3he:p}
\\
\mathcal P^n_{\hetri} &= f^{pp}_{\hetri}(E,\bm p),
\label{eq:3he:n}
\end{align}
where we consider the spectral function as a function of the separation energy $E>0$,
which is related to $\ceps$ as $\ceps=-E-\bm p^2/(4M)$ with $\bm p^2/(4M)$
as the recoil energy of the residual two-nucleon system,
and $\ceps_{32}\approx-7.72$ and $\ceps_d\approx-2.22\mev$ are the binding energies of $^3$He  and the deuteron, respectively.
Similarly, for the $^3$H nucleus, the neutron spectral function involves contributions
from the bound $pn$ state and from the $pn$ continuum states, while
the proton spectral function includes only the $nn$ continuum states:
\begin{align}
\mathcal P^n_{\htri} &= f^d_{\htri}(\bm p)\delta(E+\ceps_{31}-\ceps_d) + f^{pn}_{\htri}(E,\bm p),
\label{eq:3h:n}
\\
\mathcal P^p_{\htri} &= f^{nn}_{\htri}(E,\bm p),
\label{eq:3h:p}
\end{align}
where $\ceps_{31}\approx-8.48\mev$ is the $^3$H binding energy.

\section{QCD analysis}

We constrain the proton PDFs, the HT corrections, and the proton and the neutron
OS functions, $\delta f^p$ and $\delta f^n$,
in a global QCD analysis including the charged-lepton DIS data off $^1$H, $^2$H, $\htri$, and $\hetri$ targets,
combined with the ones from the $W^\pm/Z$ boson production at D0 and LHC experiments.
The main datasets used in our analysis are described in Refs.~\cite{Alekhin:2017fpf,Alekhin:2022tip}.
\footnote{%
We will also include the recent SeaQuest data~\protect\cite{SeaQuest:2021zxb} on the isospin asymmetry of antiquark distribution in the proton in future studies.
}
In addition, we employ the recent data on the ratio of the DIS cross sections of the three-body nuclei,
$\sigma^{\hetri}/\sigma^{\htri}$, from the \mara{} experiment~\cite{MARATHON:2021vqu}.
This allows us to study the isospin dependence of nuclear corrections, and, in particular, the neutron-proton asymmetry  $\delta f^a=\delta f^n-\delta f^p$.
To ensure a perturbative QCD description and for consistency with the previous studies~\cite{Alekhin:2017kpj,Alekhin:2022tip}, we apply the cuts  $Q^2>2.5$ and $W^2>3\gevsq$.

The point-to-point correlations in
the data are accounted in the fit whenever available.
For the \mara{} data, we combine in quadrature the published point-to-point systematic uncertainties
with the statistical (uncorrelated) ones.
We keep fixed the normalization of the most precise datasets, including the \mara{} $\sigma^{\hetri}/\sigma^{\htri}$ one,
and use them for the calibration of the other datasets (see Table~1 of Ref.~\cite{Alekhin:2022tip}).

The PDFs are parametrized following Ref.~\cite{Alekhin:2017kpj}.
The $Q^2$ dependence of the LT SFs is computed  to the next-to-next-to-leading order (NNLO) in perturbative QCD.
The functions $H_i(x)$ in the aHT model are treated independently for $i=T,2$ and are
parametrized in terms of spline polynomials interpolating between the points $x=(0, 0.1, 0.3, 0.5, 0.7, 0.9, 1)$.
A similar procedure is applied to the functions $h_i$ in the mHT model.
To reduce the number of unknown quantities in our fit, we assume $H_i^p=H_i^n$ in the aHT model.
We also test the assumption $h_i^p=h_i^n$ in the mHT model.

The nuclear effects are treated using \eq{eq:IA}.
In this approach, the nuclear corrections are driven by the momentum distribution, the nuclear binding, and the OS effect.
It was verified~\cite{Alekhin:2017fpf} that other nuclear effects,
such as the meson-exchange currents and the nuclear shadowing,
are within experimental uncertainties and therefore neglected in the present analysis.
We use the deuteron wave function computed with the Argonne potential~\cite{Wiringa:1994wb,Veerasamy:2011ak} (AV18),
and the $^3$He and $^3$H spectral functions of Ref.~\cite{Pace:2001cm}
computed with the AV18 nucleon-nucleon force and accounting for the Urbana three-nucleon interaction as well as the Coulomb effect in $\hetri$.
It was also verified that the use of the $\hetri$ spectral function of Ref.~\cite{Schulze:1992mb} and the $\htri$ spectral function obtained from isospin symmetry,
i.e., $f^d_{\htri}=f^d_{\hetri}$,  $f^{pn}_{\htri}=f^{pn}_{\hetri}$, and $f^{nn}_{\htri}=f^{pp}_{\hetri}$,
does not essentially change the results~\cite{Kulagin:2010gd}.

\begin{figure}[htb]
\centering
\hspace{-1.5em}
\includegraphics[width=1.03\linewidth]{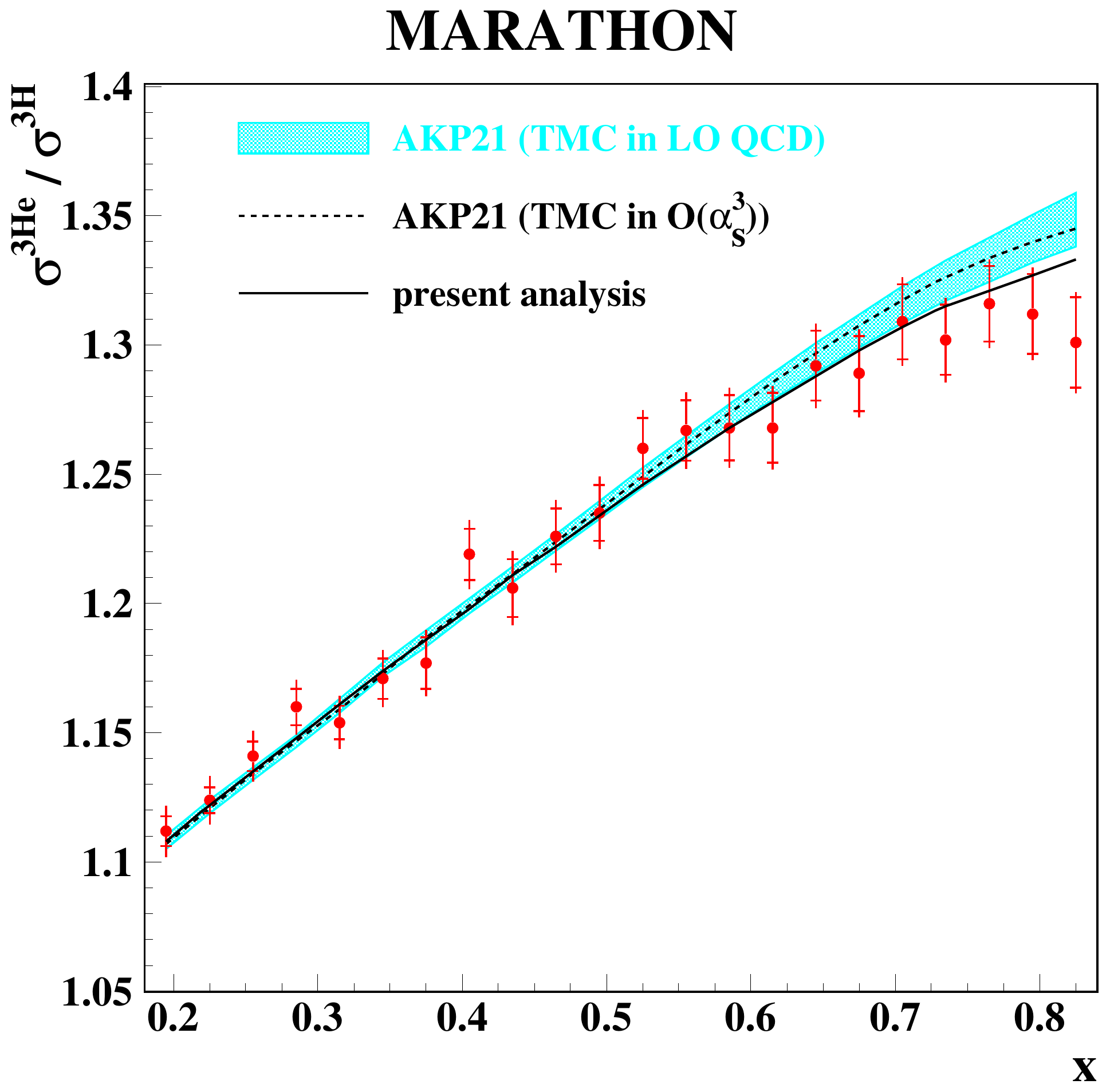}
\caption{%
Comparison of our default fit result (solid line) with
the data on the ratio of the DIS cross sections $\sigma^{\hetri}/\sigma^{\htri}$ from the \mara{} experiment~\cite{MARATHON:2021vqu}.
Also shown are the $1\sigma$ uncertainty band of analysis~\cite{Alekhin:2022tip} (AKP21) performed without the \mara{} data (shaded area) and a variant of the AKP21 predictions with
the terms up to N$^3$LO order in the QCD coupling constant~\cite{Vermaseren:2005qc} accounted for in the TMC (dashed line).}
\label{fig:fig1}
\end{figure}

Computing the nuclear SFs requires both
an energy-momentum integration and light-cone momentum integrations inside TMC and NNLO SFs.
Such integrations significantly slow down the fitting procedure.
To optimize the computing performance we treat TMC on the NNLO SFs as
$F_i^\text{TMC}=(F_i^\text{TMC}/F_i^\text{LT})^\text{LO}F_i^\text{LT(NNLO)}$,
i.e., TMC is effectively applied to the leading order (LO) SFs.
We verified that such an approximation has little impact on the predictions of Ref.~\cite{Alekhin:2022tip} for the MARATHON data.
In particular, the calculations including terms up to N$^3$LO order in the QCD coupling constant~\cite{Vermaseren:2005qc,Blumlein:2022gpp} are in  good agreement with such an approximation (Fig.~\ref{fig:fig1}).
The corresponding predictions are within $1\sigma$ of the results of Ref.~\cite{Alekhin:2022tip}, thus allowing us to safely include these data into the present fit.

The function $\delta f(x)$ is determined phenomenologically from a global fit and its $x$ dependence is parametrized as~\cite{Alekhin:2017fpf,Alekhin:2022tip}
\begin{equation}\label{eq:dfpar1}
\delta f(x)=a+bx+cx^2,
\end{equation}
where the parameters $a$, $b$, and $c$ are determined simultaneously with those of the proton PDFs and HTs.
\footnote{The correlation matrix is available upon request.}
We perform a number of fits with different setup.
In our default setup, we assume equal OS functions for the proton and neutron, $\delta f^p=\delta f^n=\delta f$, and the aHT model for the HT terms.
With such settings, we obtain a good agreement with the \mara{} data on
the ratio $\sigma^{\hetri}/\sigma^{\htri}$~\cite{MARATHON:2021vqu} with $\chi^2$ per number of data points (NDP) of $20/22$, as shown in Fig.~\ref{fig:fig1}.
Considering all data points included in our fit, we have $\chi^2/\text{NDP}=4861/4065$.
We verified that the MARATHON nuclear data do not deteriorate the description of the other datasets.
In particular, we have $\chi^2/\text{NDP}=42/31$ and 45/32 for, respectively, 7 and 8 TeV LHCb data~\cite{LHCb:2015okr,LHCb:2015kwa,LHCb:2015mad}, to be compared with the values 45/31 and 40/32 of the analysis with no nuclear data~\cite{Alekhin:2017kpj}. A small difference between the present result and Ref.~\cite{Alekhin:2017kpj} is within statistical fluctuations of data.

\begin{figure*}[hbt!]
\centering
\hspace{-1em}%
\includegraphics[width=0.5\linewidth]{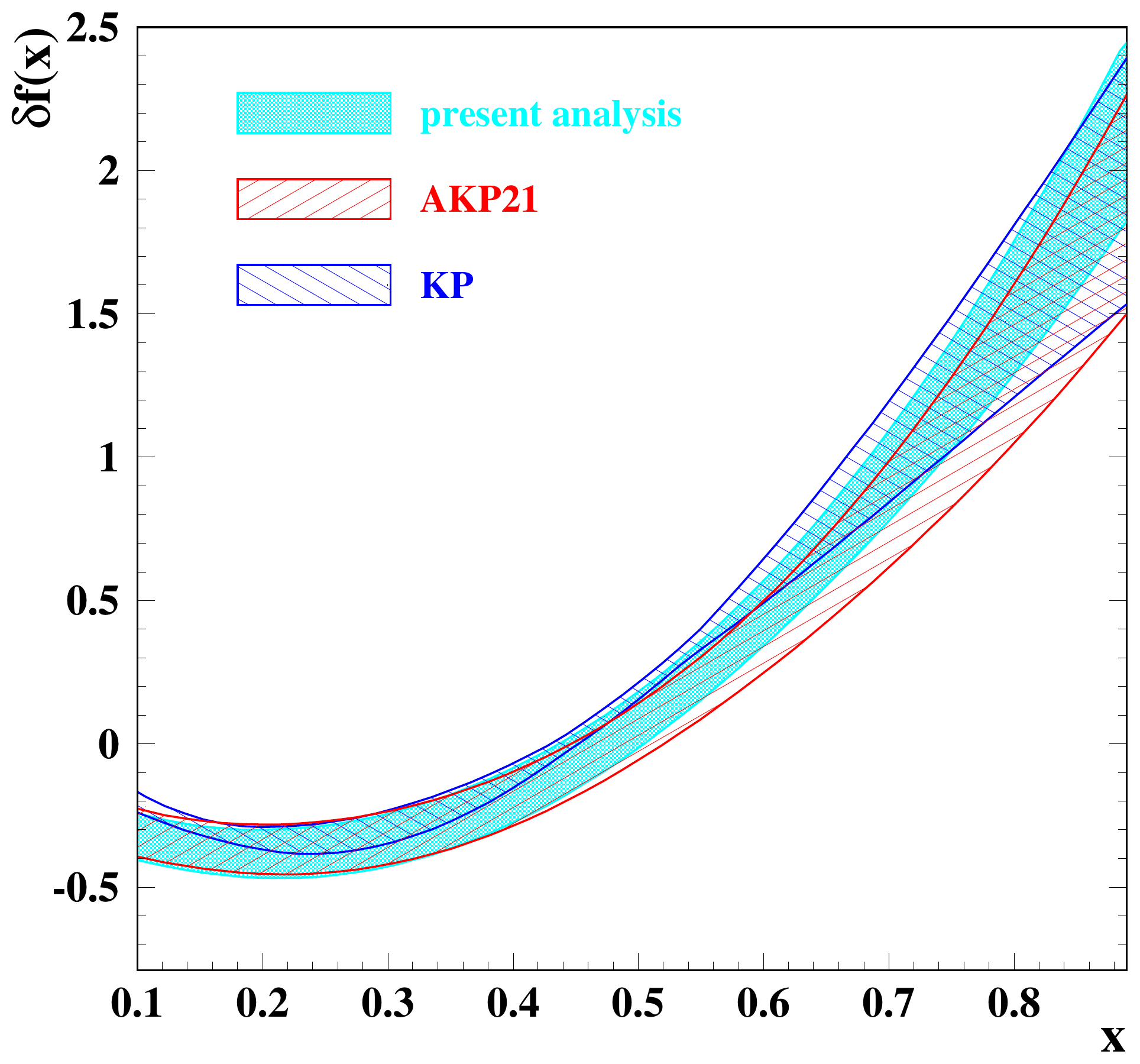}%
\hspace{1em}%
\includegraphics[width=0.5\linewidth]{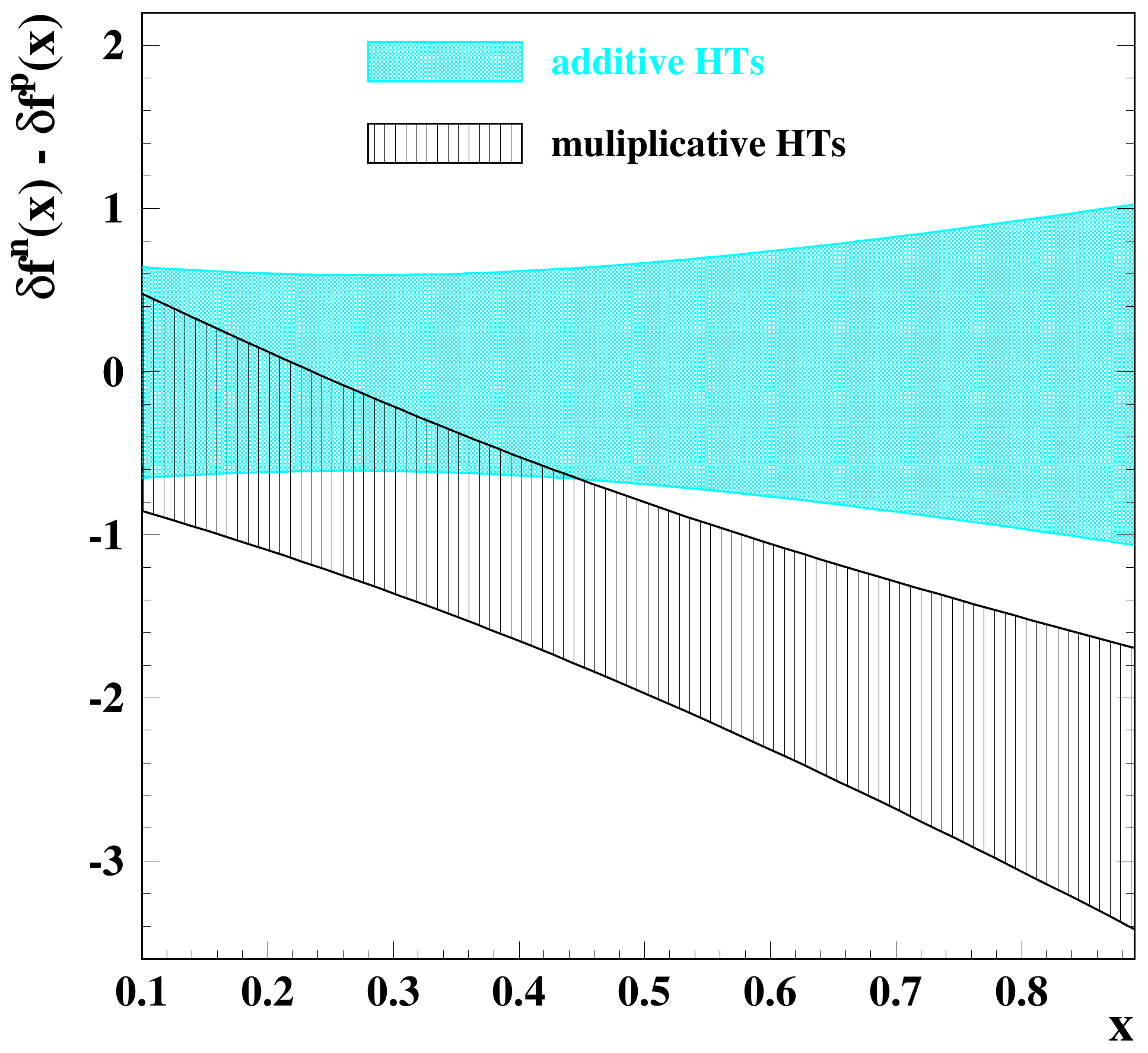}
\caption{Left:
$1\sigma$  uncertainty band on the OS function obtained assuming $\delta f^p=\delta f^n$ and the aHT model for the HT terms (shaded area).
Also shown are the results of Refs.~\cite{Kulagin:2004ie} (KP) and \cite{Alekhin:2022tip} (AKP21).
Right:
$1\sigma$ uncertainty band on the neutron-proton asymmetry $\delta f^n(x)-\delta f^p(x)$ for the aHT (shaded area) and mHT (hashed area) models.}
\label{fig:fig2}
\end{figure*}

Our results on the function $\delta f(x)$ are
shown in Fig.~\ref{fig:fig2} (left panel), together with ones from Refs.~\cite{Kulagin:2004ie,Alekhin:2022tip}.
The present results are in good agreement with
the analysis of Ref.~\cite{Kulagin:2004ie},
in which the function $\delta f(x)$ was determined
from a fit to the data on the ratios $\sigma^A/\sigma^d$ for the
DIS cross sections off nuclear targets with a mass number $A\geq 4$
using the proton and the neutron SFs of Ref.~\cite{Alekhin:2007fh}.
Our results are also in accord with the analysis of Ref.~\cite{Alekhin:2022tip},
which does not include the \mara{} data from $A=3$ nuclei.
The addition of the \mara{} data on the ratio $\sigma^{\hetri}/\sigma^{\htri}$ in the fit
allows a reduction of the $\delta f(x)$ uncertainty at large $x$.

In order to study the sensitivity of our results to the functional form of $\delta f(x)$,
we performed a fit with the term $x^3$ included in \eq{eq:dfpar1} and verified that this does not improve the fit accuracy. The KP error band in Fig.~\ref{fig:fig2} includes systematic uncertainties from the functional form as well as from the nuclear spectral function.

The results presented in Figs.~\ref{fig:fig1} and \ref{fig:fig2} (left panel) are obtained assuming an isospin
symmetric function $\delta f^p=\delta f^n$ and the aHT model for the HT terms.
The validity of this approximation was verified in the analysis of the nuclear SFs (EMC effect) in Ref.~\cite{Kulagin:2004ie}.
The same approximation was also used in  Refs.~\cite{Alekhin:2017fpf,Alekhin:2022tip}.
In this study we use the \mara{} data on the $\hetri$ and $\htri$ nuclei to constrain the asymmetry $\delta f^a=\delta f^n-\delta f^p$.
To this end, we perform a fit in which $\delta f^p$ is parametrized by \eq{eq:dfpar1},
while for the neutron-proton asymmetry we assume a linear function, $\delta f^a(x)=a_1+b_1x$.
For $\delta f^p$ we obtain a result similar to that of the isospin-symmetric fit shown in Fig.~\ref{fig:fig2} (left panel).
The corresponding asymmetry $\delta f^a$ is in a broad agreement with zero, see Fig.~\ref{fig:fig2} (right panel).

We also studied the sensitivity
of the functions $\delta f(x)$ and $\delta f^a(x)$ obtained in the fit to the underlying model for the HT terms.
In the default fit (i.e., fixed $\delta f^a=0$), we found identical $\delta f(x)$ within
uncertainties for both the aHT and the mHT models.
For this reason, we only show the results of the aHT model in the left panel of Fig.~\ref{fig:fig2}.
However, the results on the function $\delta f^a$ differ substantially in the aHT and mHT models, as shown in the right panel of Fig.~\ref{fig:fig2}.

It should be emphasized that in the mHT model the HT terms are directly correlated with the LT SFs.
As a result, the functions $H_i=F_i^\text{LT}(x,Q^2)h_i(x)$ depend on the scale $Q$, making comparisons between the aHT and mHT models sensitive to the data selection and the kinematic cuts.
The HT terms obtained in the present study are similar to those of Ref.~\cite{Alekhin:2022tip} (see Fig.~5 in \cite{Alekhin:2022tip}).
Note that the factor $F_i^\text{LT}$ introduces a nucleon isospin dependence in the HT terms even if $h_i^p=h_i^n$.
Therefore, the nonzero asymmetry $\delta f^a$ in this model (right panel of Fig.~\ref{fig:fig2}) may partially compensate the isospin dependence of the HT terms from the factor $F_i^\text{LT}$.

\begin{figure}[htb]
\centering
\hspace{-1em}%
\includegraphics[width=1.03\linewidth]{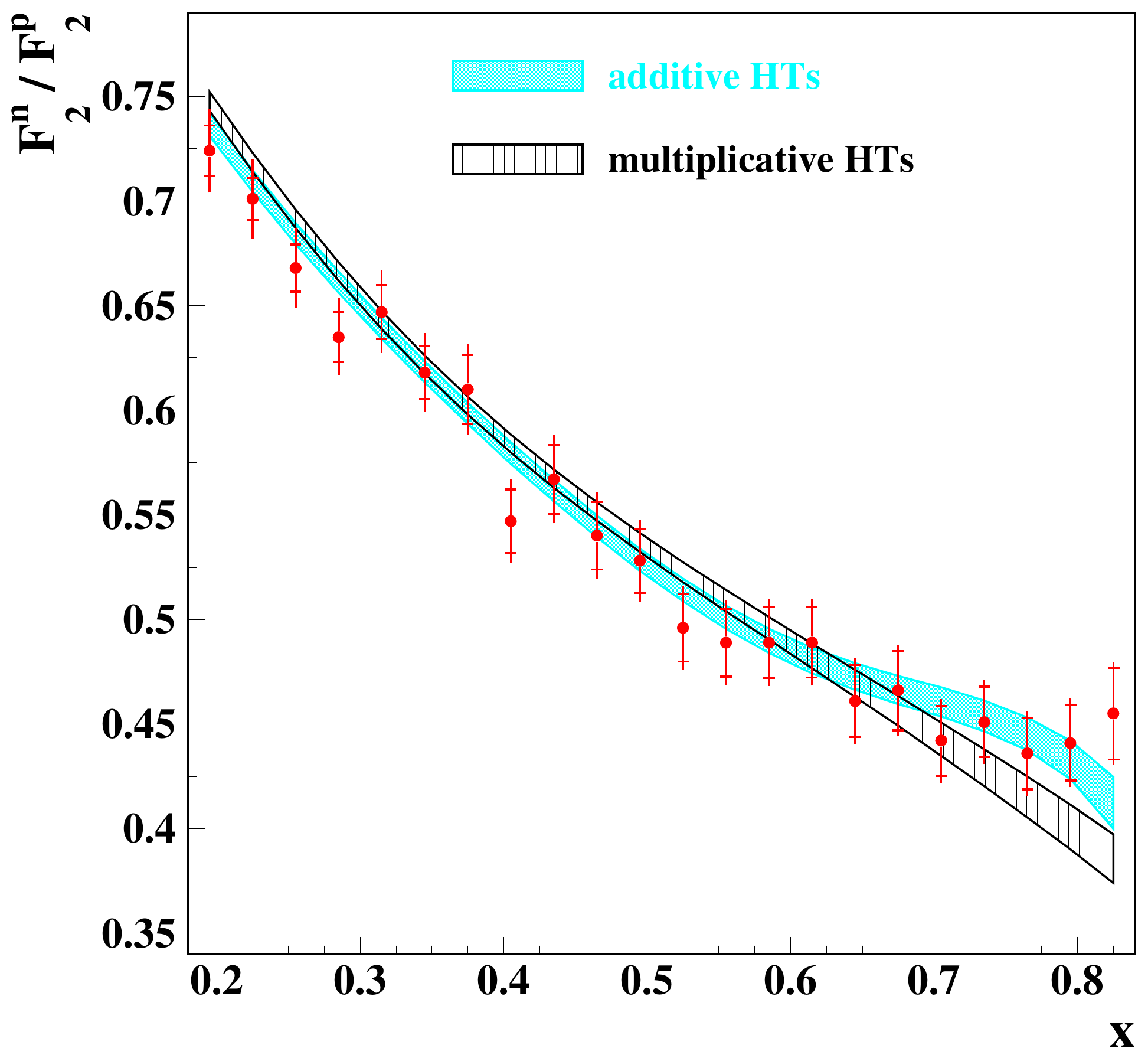}
\caption{%
Comparison of the \mara{} data on $F_2^n/F_2^p$~\cite{MARATHON:2021vqu} with
our $1\sigma$ uncertainty band predictions based on the aHT model (shaded area) and the mHT model (hashed area).}
\label{fig:fig3}
\end{figure}
\begin{figure}[htb]
\centering
\hspace{-1em}%
\includegraphics[width=1.03\linewidth]{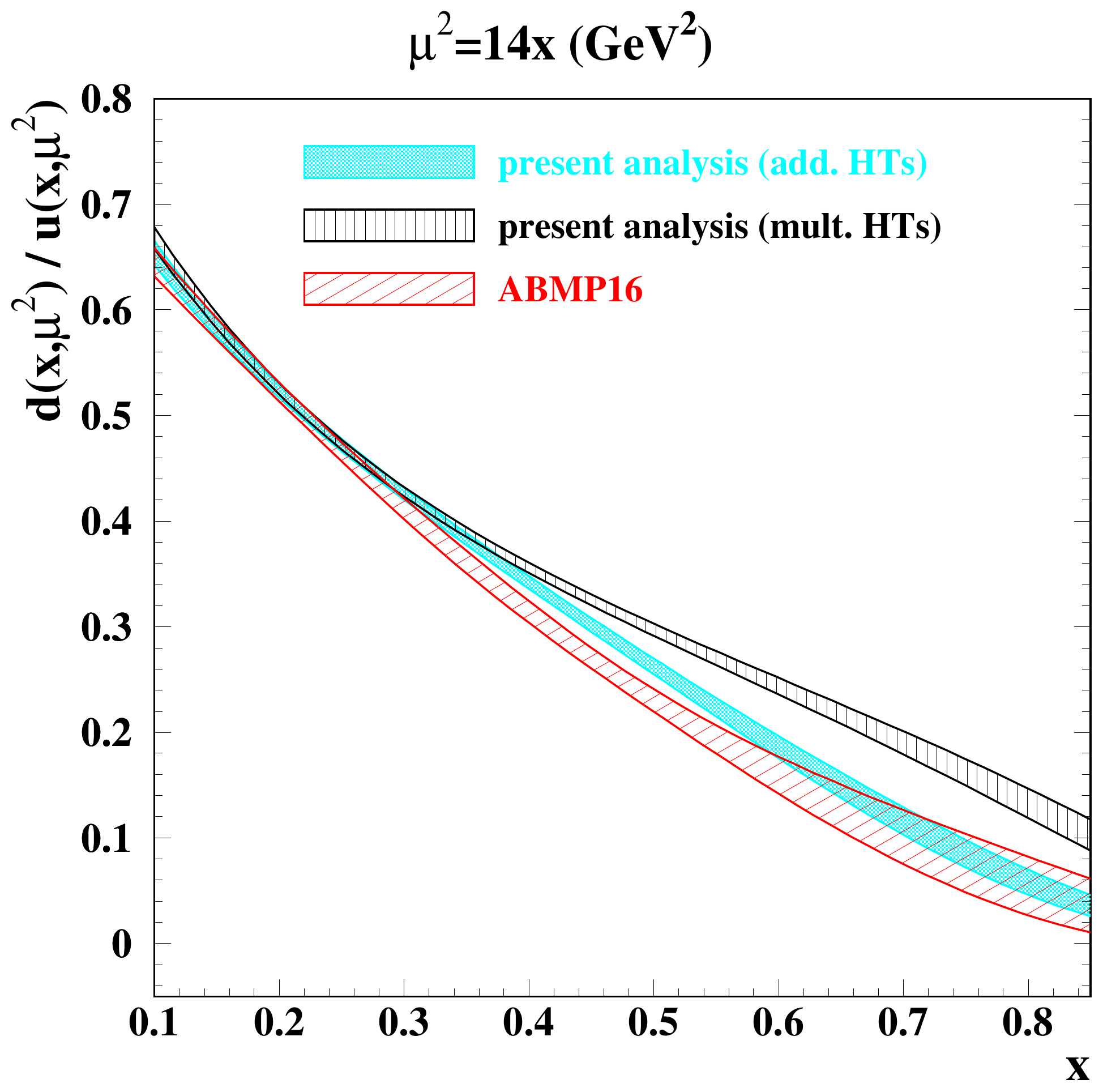}
\caption{%
Comparison of $1\sigma$ uncertainty bands for the $d/u$ ratio in the proton  obtained with the aHT (shaded area) and mHT (right-tilted hashed area) models at the factorization scale $\mu^2=14x\ (\text{GeV}^2)$.
Also shown are the results of ABMP16 analysis~\cite{Alekhin:2017kpj} (left-tilted hashed area), which does not include any nuclear data.
}\label{fig:fig4}
\vspace{-1ex}
\end{figure}

A good description of the \mara{} $F_2^n/F_2^p$ data for $x \leq 0.7$ is obtained for both the aHT and the mHT models (see Fig.~\ref{fig:fig3}).
However, for larger $x$ the aHT model provides better description of data.
The total value of $\chi^2/\text{NDP} = 20/22$ for our default fit with the aHT model to be compared with the corresponding value 34/22 of the mHT model.

It is instructive to compare the PDF ratio $d/u$ obtained with different HT models.
This comparison, shown in Fig.~\ref{fig:fig4} for the kinematics of the \mara{} experiment, indicates that the ratio $d/u$ at large $x$ is significantly higher in the mHT model.
Figure~\ref{fig:fig4} also shows the ratio $d/u$ from the analysis of Ref.~\cite{Alekhin:2015cza} (ABMP16), which was performed with the aHT model but without any nuclear data.
In this case, the ratio $d/u$ is mostly constrained by forward $W$-boson production data from the LHCb~\cite{LHCb:2015kwa,LHCb:2015okr,LHCb:2015mad} and D0~\cite{D0:2014kma} experiments.
The ABMP16 result is in good agreement with the present one obtained with the aHT model.
Instead, for the mHT model we have a significant enhancement of the ratio $d/u$  at large $x$,  which appears to be correlated with the nonzero values of the asymmetry $\delta f^a$
(cf. Figs.~\ref{fig:fig2} and \ref{fig:fig4}).
This observation demonstrates a tension in obtaining a simultaneous description of the DIS and Drell-Yan data in the mHT model.

\section{Discussion and outlook}

In summary, we obtain a good description of
data with the simple assumption of isoscalar HT contributions in the aHT model. From a QCD analysis of the precise \mara{} data on $^3$He and $^3$H mirror nuclei, we obtain the same OS function for both protons and neutrons within the uncertainties.
This nucleon OS function is
consistent with our former observations from the global QCD analyses including $^2$H DIS data~\cite{Alekhin:2017fpf,Alekhin:2022tip}, as well as with the analysis of the nuclear DIS data with
$A\geq 3$~\cite{Kulagin:2004ie,Kulagin:2010gd}.
Furthermore, the resulting $d/u$ ratio for the proton is similar to the one obtained in Ref.~\cite{Alekhin:2017kpj} without the use of any nuclear data.
The addition of DIS data from $^2$H, $^3$He, and $^3$H targets in the present QCD analysis allows a significant reduction of the uncertainty on the proton $d/u$ ratio at large $x$.

In contrast with the aHT model, in the mHT model the HT terms are different for protons and neutrons,
due to a correlation with the LT terms.
In the mHT model we find a nonzero neutron-proton asymmetry in the OS function.
The ratio $d/u$ at large $x$ is correspondingly enhanced in the mHT model as compared to that in the aHT model.
These results are driven by the \mara{} $\hetri/\htri$ data and originate from the interplay between the LT and HT terms in SFs, which is inherent to the mHT model.
We therefore conclude that this feature of the mHT model can lead to potential biases and inconsistencies.
Furthermore, the recent \mara{} data clearly prefer the aHT model over the mHT one with $\chi^2/\text{NDP}=20/22$ vs 34/22.

The interplay of the OS function with the $d/u$ ratio and the HT terms that we observe in the context of the mHT model can shed some light on the recent claim about isovector nuclear EMC effects from a global QCD analysis~\cite{Cocuzza:2021rfn}.
These results appear to be also driven by the \mara{} data on $^3$He and $^3$H within the mHT model, as discussed earlier~\cite{Alekhin:2022tip}.
In the absence of an explicit isospin dependence of the $h_i(x)$ terms, the HT contributions to the $^3$He/$^3$H ratio cancel out in the mHT model.
We therefore expect similar biases in analyses of the \mara{} $^3$He/$^3$H ratio based on the LT approximation to SFs~\cite{Segarra:2021exb}.

Future precision cross section measurements with $^2$H, $^3$H and $^3$He targets
in a wide kinematical region would further allow us to address the HT model and to constrain
the isospin dependence of nuclear effects at the parton level.
These would include future flavor sensitive DIS data
at the electron-ion collider~\cite{AbdulKhalek:2021gbh} and
from both neutrino and antineutrino charged-current interactions with hydrogen and
various nuclear targets~\cite{Petti:2019asx,Petti:2022bzt}
at the long-baseline neutrino facility~\cite{DUNE:2020ypp}.

\begin{acknowledgments}
We thank M.~V.~Garzelli and S.-O.~Moch for valuable comments,
G.~Salm\`e for providing the $\htri$ and $\hetri$ spectral functions of Ref.~\cite{Pace:2001cm},
and G.~G.~Petratos for clarifications about the \mara{} data.
S.~A. is supported by the DFG Grants No. MO 1801/5-1 and No. KN 365/14-1.
R.~P. is supported by Grant No. DE-SC0010073 from the U.S. Department of Energy.
\end{acknowledgments}

\bibliography{references}

\begin{thebibliography}{38}%
\makeatletter
\providecommand \@ifxundefined [1]{%
 \@ifx{#1\undefined}
}%
\providecommand \@ifnum [1]{%
 \ifnum #1\expandafter \@firstoftwo
 \else \expandafter \@secondoftwo
 \fi
}%
\providecommand \@ifx [1]{%
 \ifx #1\expandafter \@firstoftwo
 \else \expandafter \@secondoftwo
 \fi
}%
\providecommand \natexlab [1]{#1}%
\providecommand \enquote  [1]{``#1''}%
\providecommand \bibnamefont  [1]{#1}%
\providecommand \bibfnamefont [1]{#1}%
\providecommand \citenamefont [1]{#1}%
\providecommand \href@noop [0]{\@secondoftwo}%
\providecommand \href [0]{\begingroup \@sanitize@url \@href}%
\providecommand \@href[1]{\@@startlink{#1}\@@href}%
\providecommand \@@href[1]{\endgroup#1\@@endlink}%
\providecommand \@sanitize@url [0]{\catcode `\\12\catcode `\$12\catcode
  `\&12\catcode `\#12\catcode `\^12\catcode `\_12\catcode `\%12\relax}%
\providecommand \@@startlink[1]{}%
\providecommand \@@endlink[0]{}%
\providecommand \url  [0]{\begingroup\@sanitize@url \@url }%
\providecommand \@url [1]{\endgroup\@href {#1}{\urlprefix }}%
\providecommand \urlprefix  [0]{URL }%
\providecommand \Eprint [0]{\href }%
\providecommand \doibase [0]{https://doi.org/}%
\providecommand \selectlanguage [0]{\@gobble}%
\providecommand \bibinfo  [0]{\@secondoftwo}%
\providecommand \bibfield  [0]{\@secondoftwo}%
\providecommand \translation [1]{[#1]}%
\providecommand \BibitemOpen [0]{}%
\providecommand \bibitemStop [0]{}%
\providecommand \bibitemNoStop [0]{.\EOS\space}%
\providecommand \EOS [0]{\spacefactor3000\relax}%
\providecommand \BibitemShut  [1]{\csname bibitem#1\endcsname}%
\let\auto@bib@innerbib\@empty
\bibitem [{\citenamefont {Accardi}\ \emph {et~al.}(2016)\citenamefont {Accardi}
  \emph {et~al.}}]{Accardi:2016ndt}%
  \BibitemOpen
  \bibfield  {author} {\bibinfo {author} {\bibfnamefont {A.}~\bibnamefont
  {Accardi}} \emph {et~al.},\ }\href
  {https://doi.org/10.1140/epjc/s10052-016-4285-4} {\bibfield  {journal}
  {\bibinfo  {journal} {Eur. Phys. J. C}\ }\textbf {\bibinfo {volume} {76}},\
  \bibinfo {pages} {471} (\bibinfo {year} {2016})},\ \Eprint
  {https://arxiv.org/abs/1603.08906} {arXiv:1603.08906 [hep-ph]} \BibitemShut
  {NoStop}%
\bibitem [{\citenamefont {Gao}\ \emph {et~al.}(2018)\citenamefont {Gao},
  \citenamefont {Harland-Lang},\ and\ \citenamefont {Rojo}}]{Gao:2017yyd}%
  \BibitemOpen
  \bibfield  {author} {\bibinfo {author} {\bibfnamefont {J.}~\bibnamefont
  {Gao}}, \bibinfo {author} {\bibfnamefont {L.}~\bibnamefont {Harland-Lang}},\
  and\ \bibinfo {author} {\bibfnamefont {J.}~\bibnamefont {Rojo}},\ }\href
  {https://doi.org/10.1016/j.physrep.2018.03.002} {\bibfield  {journal}
  {\bibinfo  {journal} {Phys. Rept.}\ }\textbf {\bibinfo {volume} {742}},\
  \bibinfo {pages} {1} (\bibinfo {year} {2018})},\ \Eprint
  {https://arxiv.org/abs/1709.04922} {arXiv:1709.04922 [hep-ph]} \BibitemShut
  {NoStop}%
\bibitem [{\citenamefont {Ethier}\ and\ \citenamefont
  {Nocera}(2020)}]{Ethier:2020way}%
  \BibitemOpen
  \bibfield  {author} {\bibinfo {author} {\bibfnamefont {J.~J.}\ \bibnamefont
  {Ethier}}\ and\ \bibinfo {author} {\bibfnamefont {E.~R.}\ \bibnamefont
  {Nocera}},\ }\href {https://doi.org/10.1146/annurev-nucl-011720-042725}
  {\bibfield  {journal} {\bibinfo  {journal} {Ann. Rev. Nucl. Part. Sci.}\
  }\textbf {\bibinfo {volume} {70}},\ \bibinfo {pages} {43} (\bibinfo {year}
  {2020})},\ \Eprint {https://arxiv.org/abs/2001.07722} {arXiv:2001.07722
  [hep-ph]} \BibitemShut {NoStop}%
\bibitem [{\citenamefont {Kova\v{r}\'\i{}k}\ \emph {et~al.}(2020)\citenamefont
  {Kova\v{r}\'\i{}k}, \citenamefont {Nadolsky},\ and\ \citenamefont
  {Soper}}]{Kovarik:2019xvh}%
  \BibitemOpen
  \bibfield  {author} {\bibinfo {author} {\bibfnamefont {K.}~\bibnamefont
  {Kova\v{r}\'\i{}k}}, \bibinfo {author} {\bibfnamefont {P.~M.}\ \bibnamefont
  {Nadolsky}},\ and\ \bibinfo {author} {\bibfnamefont {D.~E.}\ \bibnamefont
  {Soper}},\ }\href {https://doi.org/10.1103/RevModPhys.92.045003} {\bibfield
  {journal} {\bibinfo  {journal} {Rev. Mod. Phys.}\ }\textbf {\bibinfo {volume}
  {92}},\ \bibinfo {pages} {045003} (\bibinfo {year} {2020})},\ \Eprint
  {https://arxiv.org/abs/1905.06957} {arXiv:1905.06957 [hep-ph]} \BibitemShut
  {NoStop}%
\bibitem [{\citenamefont {Kulagin}\ and\ \citenamefont
  {Petti}(2006)}]{Kulagin:2004ie}%
  \BibitemOpen
  \bibfield  {author} {\bibinfo {author} {\bibfnamefont {S.~A.}\ \bibnamefont
  {Kulagin}}\ and\ \bibinfo {author} {\bibfnamefont {R.}~\bibnamefont
  {Petti}},\ }\href {https://doi.org/10.1016/j.nuclphysa.2005.10.011}
  {\bibfield  {journal} {\bibinfo  {journal} {Nucl. Phys. A}\ }\textbf
  {\bibinfo {volume} {765}},\ \bibinfo {pages} {126} (\bibinfo {year}
  {2006})},\ \Eprint {https://arxiv.org/abs/hep-ph/0412425}
  {arXiv:hep-ph/0412425} \BibitemShut {NoStop}%
\bibitem [{\citenamefont {Kulagin}\ and\ \citenamefont
  {Petti}(2010)}]{Kulagin:2010gd}%
  \BibitemOpen
  \bibfield  {author} {\bibinfo {author} {\bibfnamefont {S.~A.}\ \bibnamefont
  {Kulagin}}\ and\ \bibinfo {author} {\bibfnamefont {R.}~\bibnamefont
  {Petti}},\ }\href {https://doi.org/10.1103/PhysRevC.82.054614} {\bibfield
  {journal} {\bibinfo  {journal} {Phys. Rev. C}\ }\textbf {\bibinfo {volume}
  {82}},\ \bibinfo {pages} {054614} (\bibinfo {year} {2010})},\ \Eprint
  {https://arxiv.org/abs/1004.3062} {arXiv:1004.3062 [hep-ph]} \BibitemShut
  {NoStop}%
\bibitem [{\citenamefont {Kulagin}\ and\ \citenamefont
  {Petti}(2007)}]{Kulagin:2007ju}%
  \BibitemOpen
  \bibfield  {author} {\bibinfo {author} {\bibfnamefont {S.~A.}\ \bibnamefont
  {Kulagin}}\ and\ \bibinfo {author} {\bibfnamefont {R.}~\bibnamefont
  {Petti}},\ }\href {https://doi.org/10.1103/PhysRevD.76.094023} {\bibfield
  {journal} {\bibinfo  {journal} {Phys. Rev. D}\ }\textbf {\bibinfo {volume}
  {76}},\ \bibinfo {pages} {094023} (\bibinfo {year} {2007})},\ \Eprint
  {https://arxiv.org/abs/hep-ph/0703033} {arXiv:hep-ph/0703033} \BibitemShut
  {NoStop}%
\bibitem [{\citenamefont {Kulagin}\ and\ \citenamefont
  {Petti}(2014)}]{Kulagin:2014vsa}%
  \BibitemOpen
  \bibfield  {author} {\bibinfo {author} {\bibfnamefont {S.~A.}\ \bibnamefont
  {Kulagin}}\ and\ \bibinfo {author} {\bibfnamefont {R.}~\bibnamefont
  {Petti}},\ }\href {https://doi.org/10.1103/PhysRevC.90.045204} {\bibfield
  {journal} {\bibinfo  {journal} {Phys. Rev. C}\ }\textbf {\bibinfo {volume}
  {90}},\ \bibinfo {pages} {045204} (\bibinfo {year} {2014})},\ \Eprint
  {https://arxiv.org/abs/1405.2529} {arXiv:1405.2529 [hep-ph]} \BibitemShut
  {NoStop}%
\bibitem [{\citenamefont {Ru}\ \emph {et~al.}(2016)\citenamefont {Ru},
  \citenamefont {Kulagin}, \citenamefont {Petti},\ and\ \citenamefont
  {Zhang}}]{Ru:2016wfx}%
  \BibitemOpen
  \bibfield  {author} {\bibinfo {author} {\bibfnamefont {P.}~\bibnamefont
  {Ru}}, \bibinfo {author} {\bibfnamefont {S.~A.}\ \bibnamefont {Kulagin}},
  \bibinfo {author} {\bibfnamefont {R.}~\bibnamefont {Petti}},\ and\ \bibinfo
  {author} {\bibfnamefont {B.-W.}\ \bibnamefont {Zhang}},\ }\href
  {https://doi.org/10.1103/PhysRevD.94.113013} {\bibfield  {journal} {\bibinfo
  {journal} {Phys. Rev. D}\ }\textbf {\bibinfo {volume} {94}},\ \bibinfo
  {pages} {113013} (\bibinfo {year} {2016})},\ \Eprint
  {https://arxiv.org/abs/1608.06835} {arXiv:1608.06835 [nucl-th]} \BibitemShut
  {NoStop}%
\bibitem [{\citenamefont {Abrams}\ \emph {et~al.}(2022)\citenamefont {Abrams}
  \emph {et~al.}}]{MARATHON:2021vqu}%
  \BibitemOpen
  \bibfield  {author} {\bibinfo {author} {\bibfnamefont {D.}~\bibnamefont
  {Abrams}} \emph {et~al.},\ }\href
  {https://doi.org/10.1103/PhysRevLett.128.132003} {\bibfield  {journal}
  {\bibinfo  {journal} {Phys. Rev. Lett.}\ }\textbf {\bibinfo {volume} {128}},\
  \bibinfo {pages} {132003} (\bibinfo {year} {2022})},\ \Eprint
  {https://arxiv.org/abs/2104.05850} {arXiv:2104.05850 [hep-ex]} \BibitemShut
  {NoStop}%
\bibitem [{\citenamefont {Georgi}\ and\ \citenamefont
  {Politzer}(1976)}]{Georgi:1976ve}%
  \BibitemOpen
  \bibfield  {author} {\bibinfo {author} {\bibfnamefont {H.}~\bibnamefont
  {Georgi}}\ and\ \bibinfo {author} {\bibfnamefont {H.~D.}\ \bibnamefont
  {Politzer}},\ }\href {https://doi.org/10.1103/PhysRevD.14.1829} {\bibfield
  {journal} {\bibinfo  {journal} {Phys. Rev. D}\ }\textbf {\bibinfo {volume}
  {14}},\ \bibinfo {pages} {1829} (\bibinfo {year} {1976})}\BibitemShut
  {NoStop}%
\bibitem [{\citenamefont {Virchaux}\ and\ \citenamefont
  {Milsztajn}(1992)}]{Virchaux:1991jc}%
  \BibitemOpen
  \bibfield  {author} {\bibinfo {author} {\bibfnamefont {M.}~\bibnamefont
  {Virchaux}}\ and\ \bibinfo {author} {\bibfnamefont {A.}~\bibnamefont
  {Milsztajn}},\ }\href {https://doi.org/10.1016/0370-2693(92)90527-B}
  {\bibfield  {journal} {\bibinfo  {journal} {Phys. Lett. B}\ }\textbf
  {\bibinfo {volume} {274}},\ \bibinfo {pages} {221} (\bibinfo {year}
  {1992})}\BibitemShut {NoStop}%
\bibitem [{\citenamefont {Akulinichev}\ \emph {et~al.}(1985)\citenamefont
  {Akulinichev}, \citenamefont {Kulagin},\ and\ \citenamefont
  {Vagradov}}]{Akulinichev:1985ij}%
  \BibitemOpen
  \bibfield  {author} {\bibinfo {author} {\bibfnamefont {S.~V.}\ \bibnamefont
  {Akulinichev}}, \bibinfo {author} {\bibfnamefont {S.~A.}\ \bibnamefont
  {Kulagin}},\ and\ \bibinfo {author} {\bibfnamefont {G.~M.}\ \bibnamefont
  {Vagradov}},\ }\href {https://doi.org/10.1016/0370-2693(85)90799-3}
  {\bibfield  {journal} {\bibinfo  {journal} {Phys. Lett. B}\ }\textbf
  {\bibinfo {volume} {158}},\ \bibinfo {pages} {485} (\bibinfo {year}
  {1985})}\BibitemShut {NoStop}%
\bibitem [{\citenamefont {Kulagin}(1989)}]{Kulagin:1989mu}%
  \BibitemOpen
  \bibfield  {author} {\bibinfo {author} {\bibfnamefont {S.~A.}\ \bibnamefont
  {Kulagin}},\ }\href {https://doi.org/10.1016/0375-9474(89)90233-9} {\bibfield
   {journal} {\bibinfo  {journal} {Nucl. Phys. A}\ }\textbf {\bibinfo {volume}
  {500}},\ \bibinfo {pages} {653} (\bibinfo {year} {1989})}\BibitemShut
  {NoStop}%
\bibitem [{\citenamefont {Kulagin}\ \emph {et~al.}(1994)\citenamefont
  {Kulagin}, \citenamefont {Piller},\ and\ \citenamefont
  {Weise}}]{Kulagin:1994fz}%
  \BibitemOpen
  \bibfield  {author} {\bibinfo {author} {\bibfnamefont {S.~A.}\ \bibnamefont
  {Kulagin}}, \bibinfo {author} {\bibfnamefont {G.}~\bibnamefont {Piller}},\
  and\ \bibinfo {author} {\bibfnamefont {W.}~\bibnamefont {Weise}},\ }\href
  {https://doi.org/10.1103/PhysRevC.50.1154} {\bibfield  {journal} {\bibinfo
  {journal} {Phys. Rev. C}\ }\textbf {\bibinfo {volume} {50}},\ \bibinfo
  {pages} {1154} (\bibinfo {year} {1994})},\ \Eprint
  {https://arxiv.org/abs/nucl-th/9402015} {arXiv:nucl-th/9402015} \BibitemShut
  {NoStop}%
\bibitem [{\citenamefont {Alekhin}\ \emph {et~al.}(2022)\citenamefont
  {Alekhin}, \citenamefont {Kulagin},\ and\ \citenamefont
  {Petti}}]{Alekhin:2022tip}%
  \BibitemOpen
  \bibfield  {author} {\bibinfo {author} {\bibfnamefont {S.~I.}\ \bibnamefont
  {Alekhin}}, \bibinfo {author} {\bibfnamefont {S.~A.}\ \bibnamefont
  {Kulagin}},\ and\ \bibinfo {author} {\bibfnamefont {R.}~\bibnamefont
  {Petti}},\ }\href {https://doi.org/10.1103/PhysRevD.105.114037} {\bibfield
  {journal} {\bibinfo  {journal} {Phys. Rev. D}\ }\textbf {\bibinfo {volume}
  {105}},\ \bibinfo {pages} {114037} (\bibinfo {year} {2022})},\ \Eprint
  {https://arxiv.org/abs/2203.07333} {arXiv:2203.07333} \BibitemShut {NoStop}%
\bibitem [{\citenamefont {Alekhin}\ \emph
  {et~al.}(2017{\natexlab{a}})\citenamefont {Alekhin}, \citenamefont
  {Kulagin},\ and\ \citenamefont {Petti}}]{Alekhin:2017fpf}%
  \BibitemOpen
  \bibfield  {author} {\bibinfo {author} {\bibfnamefont {S.~I.}\ \bibnamefont
  {Alekhin}}, \bibinfo {author} {\bibfnamefont {S.~A.}\ \bibnamefont
  {Kulagin}},\ and\ \bibinfo {author} {\bibfnamefont {R.}~\bibnamefont
  {Petti}},\ }\href {https://doi.org/10.1103/PhysRevD.96.054005} {\bibfield
  {journal} {\bibinfo  {journal} {Phys. Rev. D}\ }\textbf {\bibinfo {volume}
  {96}},\ \bibinfo {pages} {054005} (\bibinfo {year} {2017}{\natexlab{a}})},\
  \Eprint {https://arxiv.org/abs/1704.00204} {arXiv:1704.00204 [nucl-th]}
  \BibitemShut {NoStop}%
\bibitem [{\citenamefont {Kulagin}\ and\ \citenamefont
  {Melnitchouk}(2008)}]{Kulagin:2008fm}%
  \BibitemOpen
  \bibfield  {author} {\bibinfo {author} {\bibfnamefont {S.~A.}\ \bibnamefont
  {Kulagin}}\ and\ \bibinfo {author} {\bibfnamefont {W.}~\bibnamefont
  {Melnitchouk}},\ }\href {https://doi.org/10.1103/PhysRevC.78.065203}
  {\bibfield  {journal} {\bibinfo  {journal} {Phys. Rev. C}\ }\textbf {\bibinfo
  {volume} {78}},\ \bibinfo {pages} {065203} (\bibinfo {year} {2008})},\
  \Eprint {https://arxiv.org/abs/0809.3998} {arXiv:0809.3998 [nucl-th]}
  \BibitemShut {NoStop}%
\bibitem [{\citenamefont {Dove}\ \emph {et~al.}(2021)\citenamefont {Dove} \emph
  {et~al.}}]{SeaQuest:2021zxb}%
  \BibitemOpen
  \bibfield  {author} {\bibinfo {author} {\bibfnamefont {J.}~\bibnamefont
  {Dove}} \emph {et~al.} (\bibinfo {collaboration} {SeaQuest}),\ }\href
  {https://doi.org/10.1038/s41586-022-04707-z} {\bibfield  {journal} {\bibinfo
  {journal} {Nature}\ }\textbf {\bibinfo {volume} {590}},\ \bibinfo {pages}
  {561} (\bibinfo {year} {2021})},\ \bibinfo {note} {[Erratum: Nature 604, E26
  (2022)]},\ \Eprint {https://arxiv.org/abs/2103.04024} {arXiv:2103.04024
  [hep-ph]} \BibitemShut {NoStop}%
\bibitem [{\citenamefont {Alekhin}\ \emph
  {et~al.}(2017{\natexlab{b}})\citenamefont {Alekhin}, \citenamefont
  {Bl\"umlein}, \citenamefont {Moch},\ and\ \citenamefont
  {Placakyte}}]{Alekhin:2017kpj}%
  \BibitemOpen
  \bibfield  {author} {\bibinfo {author} {\bibfnamefont {S.}~\bibnamefont
  {Alekhin}}, \bibinfo {author} {\bibfnamefont {J.}~\bibnamefont {Bl\"umlein}},
  \bibinfo {author} {\bibfnamefont {S.}~\bibnamefont {Moch}},\ and\ \bibinfo
  {author} {\bibfnamefont {R.}~\bibnamefont {Placakyte}},\ }\href
  {https://doi.org/10.1103/PhysRevD.96.014011} {\bibfield  {journal} {\bibinfo
  {journal} {Phys. Rev. D}\ }\textbf {\bibinfo {volume} {96}},\ \bibinfo
  {pages} {014011} (\bibinfo {year} {2017}{\natexlab{b}})},\ \Eprint
  {https://arxiv.org/abs/1701.05838} {arXiv:1701.05838 [hep-ph]} \BibitemShut
  {NoStop}%
\bibitem [{\citenamefont {Wiringa}\ \emph {et~al.}(1995)\citenamefont
  {Wiringa}, \citenamefont {Stoks},\ and\ \citenamefont
  {Schiavilla}}]{Wiringa:1994wb}%
  \BibitemOpen
  \bibfield  {author} {\bibinfo {author} {\bibfnamefont {R.~B.}\ \bibnamefont
  {Wiringa}}, \bibinfo {author} {\bibfnamefont {V.~G.~J.}\ \bibnamefont
  {Stoks}},\ and\ \bibinfo {author} {\bibfnamefont {R.}~\bibnamefont
  {Schiavilla}},\ }\href {https://doi.org/10.1103/PhysRevC.51.38} {\bibfield
  {journal} {\bibinfo  {journal} {Phys. Rev. C}\ }\textbf {\bibinfo {volume}
  {51}},\ \bibinfo {pages} {38} (\bibinfo {year} {1995})},\ \Eprint
  {https://arxiv.org/abs/nucl-th/9408016} {arXiv:nucl-th/9408016 [nucl-th]}
  \BibitemShut {NoStop}%
\bibitem [{\citenamefont {Veerasamy}\ and\ \citenamefont
  {Polyzou}(2011)}]{Veerasamy:2011ak}%
  \BibitemOpen
  \bibfield  {author} {\bibinfo {author} {\bibfnamefont {S.}~\bibnamefont
  {Veerasamy}}\ and\ \bibinfo {author} {\bibfnamefont {W.~N.}\ \bibnamefont
  {Polyzou}},\ }\href {https://doi.org/10.1103/PhysRevC.84.034003} {\bibfield
  {journal} {\bibinfo  {journal} {Phys. Rev. C}\ }\textbf {\bibinfo {volume}
  {84}},\ \bibinfo {pages} {034003} (\bibinfo {year} {2011})},\ \Eprint
  {https://arxiv.org/abs/1106.1934} {arXiv:1106.1934 [nucl-th]} \BibitemShut
  {NoStop}%
\bibitem [{\citenamefont {Pace}\ \emph {et~al.}(2001)\citenamefont {Pace},
  \citenamefont {Salme}, \citenamefont {Scopetta},\ and\ \citenamefont
  {Kievsky}}]{Pace:2001cm}%
  \BibitemOpen
  \bibfield  {author} {\bibinfo {author} {\bibfnamefont {E.}~\bibnamefont
  {Pace}}, \bibinfo {author} {\bibfnamefont {G.}~\bibnamefont {Salme}},
  \bibinfo {author} {\bibfnamefont {S.}~\bibnamefont {Scopetta}},\ and\
  \bibinfo {author} {\bibfnamefont {A.}~\bibnamefont {Kievsky}},\ }\href
  {https://doi.org/10.1103/PhysRevC.64.055203} {\bibfield  {journal} {\bibinfo
  {journal} {Phys. Rev. C}\ }\textbf {\bibinfo {volume} {64}},\ \bibinfo
  {pages} {055203} (\bibinfo {year} {2001})},\ \Eprint
  {https://arxiv.org/abs/nucl-th/0109005} {arXiv:nucl-th/0109005} \BibitemShut
  {NoStop}%
\bibitem [{\citenamefont {Schulze}\ and\ \citenamefont
  {Sauer}(1993)}]{Schulze:1992mb}%
  \BibitemOpen
  \bibfield  {author} {\bibinfo {author} {\bibfnamefont {R.~W.}\ \bibnamefont
  {Schulze}}\ and\ \bibinfo {author} {\bibfnamefont {P.~U.}\ \bibnamefont
  {Sauer}},\ }\href {https://doi.org/10.1103/PhysRevC.48.38} {\bibfield
  {journal} {\bibinfo  {journal} {Phys. Rev. C}\ }\textbf {\bibinfo {volume}
  {48}},\ \bibinfo {pages} {38} (\bibinfo {year} {1993})}\BibitemShut {NoStop}%
\bibitem [{\citenamefont {Vermaseren}\ \emph {et~al.}(2005)\citenamefont
  {Vermaseren}, \citenamefont {Vogt},\ and\ \citenamefont
  {Moch}}]{Vermaseren:2005qc}%
  \BibitemOpen
  \bibfield  {author} {\bibinfo {author} {\bibfnamefont {J.~A.~M.}\
  \bibnamefont {Vermaseren}}, \bibinfo {author} {\bibfnamefont
  {A.}~\bibnamefont {Vogt}},\ and\ \bibinfo {author} {\bibfnamefont
  {S.}~\bibnamefont {Moch}},\ }\href
  {https://doi.org/10.1016/j.nuclphysb.2005.06.020} {\bibfield  {journal}
  {\bibinfo  {journal} {Nucl. Phys. B}\ }\textbf {\bibinfo {volume} {724}},\
  \bibinfo {pages} {3} (\bibinfo {year} {2005})},\ \Eprint
  {https://arxiv.org/abs/hep-ph/0504242} {arXiv:hep-ph/0504242} \BibitemShut
  {NoStop}%
\bibitem [{\citenamefont {Bl\"umlein}\ \emph {et~al.}(2022)\citenamefont
  {Bl\"umlein}, \citenamefont {Marquard}, \citenamefont {Schneider},\ and\
  \citenamefont {Sch\"onwald}}]{Blumlein:2022gpp}%
  \BibitemOpen
  \bibfield  {author} {\bibinfo {author} {\bibfnamefont {J.}~\bibnamefont
  {Bl\"umlein}}, \bibinfo {author} {\bibfnamefont {P.}~\bibnamefont
  {Marquard}}, \bibinfo {author} {\bibfnamefont {C.}~\bibnamefont
  {Schneider}},\ and\ \bibinfo {author} {\bibfnamefont {K.}~\bibnamefont
  {Sch\"onwald}},\ }\href {https://doi.org/10.1007/JHEP11(2022)156} {\bibfield
  {journal} {\bibinfo  {journal} {J. High Energy Phys.}\ }\textbf {\bibinfo
  {volume} {2022}}\bibfield  {number} {\bibinfo  {number} { (11)},\ \bibinfo
  {pages} {156}},\ }\Eprint {https://arxiv.org/abs/2208.14325}
  {arXiv:2208.14325 [hep-ph]} \BibitemShut {NoStop}%
\bibitem [{\citenamefont {Aaij}\ \emph
  {et~al.}(2015{\natexlab{a}})\citenamefont {Aaij} \emph
  {et~al.}}]{LHCb:2015okr}%
  \BibitemOpen
  \bibfield  {author} {\bibinfo {author} {\bibfnamefont {R.}~\bibnamefont
  {Aaij}} \emph {et~al.} (\bibinfo {collaboration} {LHCb}),\ }\href
  {https://doi.org/10.1007/JHEP08(2015)039} {\bibfield  {journal} {\bibinfo
  {journal} {J. High Energy Phys.}\ }\textbf {\bibinfo {volume}
  {2015}}\bibfield  {number} {\bibinfo  {number} { (08)},\ \bibinfo {pages}
  {039}},\ }\Eprint {https://arxiv.org/abs/1505.07024} {arXiv:1505.07024
  [hep-ex]} \BibitemShut {NoStop}%
\bibitem [{\citenamefont {Aaij}\ \emph
  {et~al.}(2015{\natexlab{b}})\citenamefont {Aaij} \emph
  {et~al.}}]{LHCb:2015kwa}%
  \BibitemOpen
  \bibfield  {author} {\bibinfo {author} {\bibfnamefont {R.}~\bibnamefont
  {Aaij}} \emph {et~al.} (\bibinfo {collaboration} {LHCb}),\ }\href
  {https://doi.org/10.1007/JHEP05(2015)109} {\bibfield  {journal} {\bibinfo
  {journal} {J. High Energy Phys.}\ }\textbf {\bibinfo {volume}
  {2015}}\bibfield  {number} {\bibinfo  {number} { (05)},\ \bibinfo {pages}
  {109}},\ }\Eprint {https://arxiv.org/abs/1503.00963} {arXiv:1503.00963
  [hep-ex]} \BibitemShut {NoStop}%
\bibitem [{\citenamefont {Aaij}\ \emph {et~al.}(2016)\citenamefont {Aaij} \emph
  {et~al.}}]{LHCb:2015mad}%
  \BibitemOpen
  \bibfield  {author} {\bibinfo {author} {\bibfnamefont {R.}~\bibnamefont
  {Aaij}} \emph {et~al.} (\bibinfo {collaboration} {LHCb}),\ }\href
  {https://doi.org/10.1007/JHEP01(2016)155} {\bibfield  {journal} {\bibinfo
  {journal} {J. High Energy Phys.}\ }\textbf {\bibinfo {volume}
  {2016}}\bibfield  {number} {\bibinfo  {number} { (01)},\ \bibinfo {pages}
  {155}},\ }\Eprint {https://arxiv.org/abs/1511.08039} {arXiv:1511.08039
  [hep-ex]} \BibitemShut {NoStop}%
\bibitem [{\citenamefont {Alekhin}\ \emph {et~al.}(2007)\citenamefont
  {Alekhin}, \citenamefont {Kulagin},\ and\ \citenamefont
  {Petti}}]{Alekhin:2007fh}%
  \BibitemOpen
  \bibfield  {author} {\bibinfo {author} {\bibfnamefont {S.}~\bibnamefont
  {Alekhin}}, \bibinfo {author} {\bibfnamefont {S.~A.}\ \bibnamefont
  {Kulagin}},\ and\ \bibinfo {author} {\bibfnamefont {R.}~\bibnamefont
  {Petti}},\ }\href {https://doi.org/10.1063/1.2834481} {\bibfield  {journal}
  {\bibinfo  {journal} {AIP Conf. Proc.}\ }\textbf {\bibinfo {volume} {967}},\
  \bibinfo {pages} {215} (\bibinfo {year} {2007})},\ \Eprint
  {https://arxiv.org/abs/0710.0124} {arXiv:0710.0124 [hep-ph]} \BibitemShut
  {NoStop}%
\bibitem [{\citenamefont {Alekhin}\ \emph {et~al.}(2016)\citenamefont
  {Alekhin}, \citenamefont {Bl\"umlein}, \citenamefont {Moch},\ and\
  \citenamefont {Pla\v{c}akyt\.{e}}}]{Alekhin:2015cza}%
  \BibitemOpen
  \bibfield  {author} {\bibinfo {author} {\bibfnamefont {S.}~\bibnamefont
  {Alekhin}}, \bibinfo {author} {\bibfnamefont {J.}~\bibnamefont {Bl\"umlein}},
  \bibinfo {author} {\bibfnamefont {S.}~\bibnamefont {Moch}},\ and\ \bibinfo
  {author} {\bibfnamefont {R.}~\bibnamefont {Pla\v{c}akyt\.{e}}},\ }\href
  {https://doi.org/10.1103/PhysRevD.94.114038} {\bibfield  {journal} {\bibinfo
  {journal} {Phys. Rev. D}\ }\textbf {\bibinfo {volume} {94}},\ \bibinfo
  {pages} {114038} (\bibinfo {year} {2016})},\ \Eprint
  {https://arxiv.org/abs/1508.07923} {arXiv:1508.07923 [hep-ph]} \BibitemShut
  {NoStop}%
\bibitem [{\citenamefont {Abazov}\ \emph {et~al.}(2015)\citenamefont {Abazov}
  \emph {et~al.}}]{D0:2014kma}%
  \BibitemOpen
  \bibfield  {author} {\bibinfo {author} {\bibfnamefont {V.~M.}\ \bibnamefont
  {Abazov}} \emph {et~al.} (\bibinfo {collaboration} {D0}),\ }\href
  {https://doi.org/10.1103/PhysRevD.91.032007} {\bibfield  {journal} {\bibinfo
  {journal} {Phys. Rev. D}\ }\textbf {\bibinfo {volume} {91}},\ \bibinfo
  {pages} {032007} (\bibinfo {year} {2015})},\ \bibinfo {note} {[Erratum: Phys.
  Rev. D 91, 079901 (2015)]},\ \Eprint {https://arxiv.org/abs/1412.2862}
  {arXiv:1412.2862 [hep-ex]} \BibitemShut {NoStop}%
\bibitem [{\citenamefont {Cocuzza}\ \emph {et~al.}(2021)\citenamefont
  {Cocuzza}, \citenamefont {Keppel}, \citenamefont {Liu}, \citenamefont
  {Melnitchouk}, \citenamefont {Metz}, \citenamefont {Sato},\ and\
  \citenamefont {Thomas}}]{Cocuzza:2021rfn}%
  \BibitemOpen
  \bibfield  {author} {\bibinfo {author} {\bibfnamefont {C.}~\bibnamefont
  {Cocuzza}}, \bibinfo {author} {\bibfnamefont {C.~E.}\ \bibnamefont {Keppel}},
  \bibinfo {author} {\bibfnamefont {H.}~\bibnamefont {Liu}}, \bibinfo {author}
  {\bibfnamefont {W.}~\bibnamefont {Melnitchouk}}, \bibinfo {author}
  {\bibfnamefont {A.}~\bibnamefont {Metz}}, \bibinfo {author} {\bibfnamefont
  {N.}~\bibnamefont {Sato}},\ and\ \bibinfo {author} {\bibfnamefont {A.~W.}\
  \bibnamefont {Thomas}},\ }\href
  {https://doi.org/10.1103/PhysRevLett.127.242001} {\bibfield  {journal}
  {\bibinfo  {journal} {Phys. Rev. Lett.}\ }\textbf {\bibinfo {volume} {127}},\
  \bibinfo {pages} {242001} (\bibinfo {year} {2021})},\ \Eprint
  {https://arxiv.org/abs/2104.06946} {arXiv:2104.06946 [hep-ph]} \BibitemShut
  {NoStop}%
\bibitem [{\citenamefont {Segarra}\ \emph {et~al.}(2021)\citenamefont {Segarra}
  \emph {et~al.}}]{Segarra:2021exb}%
  \BibitemOpen
  \bibfield  {author} {\bibinfo {author} {\bibfnamefont {E.~P.}\ \bibnamefont
  {Segarra}} \emph {et~al.},\ }\Eprint {https://arxiv.org/abs/2104.07130}
  {arXiv:2104.07130 [hep-ph]}  (\bibinfo {year} {2021})\BibitemShut {NoStop}%
\bibitem [{\citenamefont {Abdul~Khalek}\ \emph {et~al.}(2022)\citenamefont
  {Abdul~Khalek} \emph {et~al.}}]{AbdulKhalek:2021gbh}%
  \BibitemOpen
  \bibfield  {author} {\bibinfo {author} {\bibfnamefont {R.}~\bibnamefont
  {Abdul~Khalek}} \emph {et~al.},\ }\href
  {https://doi.org/10.1016/j.nuclphysa.2022.122447} {\bibfield  {journal}
  {\bibinfo  {journal} {Nucl. Phys. A}\ }\textbf {\bibinfo {volume} {1026}},\
  \bibinfo {pages} {122447} (\bibinfo {year} {2022})},\ \Eprint
  {https://arxiv.org/abs/2103.05419} {arXiv:2103.05419 [physics.ins-det]}
  \BibitemShut {NoStop}%
\bibitem [{\citenamefont {Petti}(2019)}]{Petti:2019asx}%
  \BibitemOpen
  \bibfield  {author} {\bibinfo {author} {\bibfnamefont {R.}~\bibnamefont
  {Petti}},\ }\Eprint {https://arxiv.org/abs/1910.05995} {arXiv:1910.05995
  [hep-ex]}  (\bibinfo {year} {2019})\BibitemShut {NoStop}%
\bibitem [{\citenamefont {Petti}(2022)}]{Petti:2022bzt}%
  \BibitemOpen
  \bibfield  {author} {\bibinfo {author} {\bibfnamefont {R.}~\bibnamefont
  {Petti}},\ }\href {https://doi.org/10.1016/j.physletb.2022.137469} {\bibfield
   {journal} {\bibinfo  {journal} {Phys. Lett. B}\ }\textbf {\bibinfo {volume}
  {834}},\ \bibinfo {pages} {137469} (\bibinfo {year} {2022})},\ \Eprint
  {https://arxiv.org/abs/2205.10396} {arXiv:2205.10396 [hep-ph]} \BibitemShut
  {NoStop}%
\bibitem [{\citenamefont {Abi}\ \emph {et~al.}(2020)\citenamefont {Abi} \emph
  {et~al.}}]{DUNE:2020ypp}%
  \BibitemOpen
  \bibfield  {author} {\bibinfo {author} {\bibfnamefont {B.}~\bibnamefont
  {Abi}} \emph {et~al.} (\bibinfo {collaboration} {DUNE}),\ }\Eprint
  {https://arxiv.org/abs/2002.03005} {arXiv:2002.03005 [hep-ex]}  (\bibinfo
  {year} {2020})\BibitemShut {NoStop}%
\end{thebibliography}%
\end{document}